\documentclass[journal]{IEEEtran}
\IEEEoverridecommandlockouts
\usepackage{lineno}
\usepackage{cite}
\usepackage{hyperref}
\usepackage{amsmath,amssymb,amsfonts}
\usepackage{amsmath}
\usepackage{amsthm}

\usepackage{graphicx}
\usepackage{textcomp}
\usepackage{xcolor}
\usepackage{graphicx}
\usepackage{float}
\usepackage{subfigure}
\usepackage{amsmath}
\usepackage{amsfonts,amssymb}
\usepackage{mathrsfs}
\usepackage{mathtools}
\usepackage{algorithm}
\usepackage{algorithmicx}
\usepackage{algpseudocode}
\usepackage{bm}
\usepackage{multirow}
\usepackage{array}
\usepackage{amssymb}
\usepackage{amsmath}
\usepackage{cite}
\usepackage{url}
\usepackage{xcolor}
\usepackage{cite,graphicx,amsmath,amssymb}
\usepackage{subfigure}
\usepackage{fancyhdr}
\usepackage{mdwmath}
\usepackage{mdwtab}
\usepackage{caption}
\usepackage{amsthm}
\usepackage{setspace}
\usepackage{bm}
\usepackage{algorithm}
\usepackage{algpseudocode}
\usepackage{mathtools}
\usepackage{dsfont}
\usepackage{bbm}
\usepackage{cases}
\newtheorem{remark}{Remark}
\theoremstyle{definition}
\newtheorem{theorem}{Theorem}

\newtheorem{lemma}{Lemma}

\newtheorem{corollary}{Corollary}

\makeatletter
\newcommand{\biggg}{\bBigg@{3}}
\newcommand{\Biggg}{\bBigg@{3.5}}
\makeatother
\begin{document}
\title{On the Impact of Reactive Region on the Near-Field Channel Gain}
\author{Chongjun~Ouyang, Zhaolin~Wang, Boqun~Zhao, Xingqi~Zhang, and Yuanwei~Liu
\thanks{C. Ouyang is with University College Dublin, Dublin, D04 V1W8, Ireland, and also with Queen Mary University of London, London, E1 4NS, U.K. (e-mail: chongjun.ouyang@ucd.ie).}
\thanks{Z. Wang and Y. Liu are with the Department of Electronic Engineering and Computer Science, Queen Mary University of London, London, E1 4NS, U.K. (e-mail: \{zhaolin.wang, yuanwei.liu\}@qmul.ac.uk).}
\thanks{B. Zhao and X. Zhang are with the Department of Electrical and Computer Engineering, University of Alberta, Edmonton, AB T6G 1H9, Canada (email: \{boqun1, xingqi.zhang\}@ualberta.ca).}}
\maketitle
\begin{abstract}
The near-field channel gain is analyzed by considering both radiating and reactive components of the electromagnetic field. Novel expressions are derived for the channel gains of spatially-discrete (SPD) and continuous-aperture (CAP) arrays, which are more accurate than conventional results that neglect the reactive region. To gain further insights, asymptotic analyses are carried out in the large aperture size, based on which the impact of the reactive region is discussed. It is proved that for both SPD and CAP arrays, the impact of the reactive region on near-field channel gain is negligible, even as the array aperture size approaches infinity. 
\end{abstract} 
\section{Introduction}
\begin{figure}[!t]
 \centering
\setlength{\abovecaptionskip}{0pt}
\includegraphics[height=0.15\textwidth]{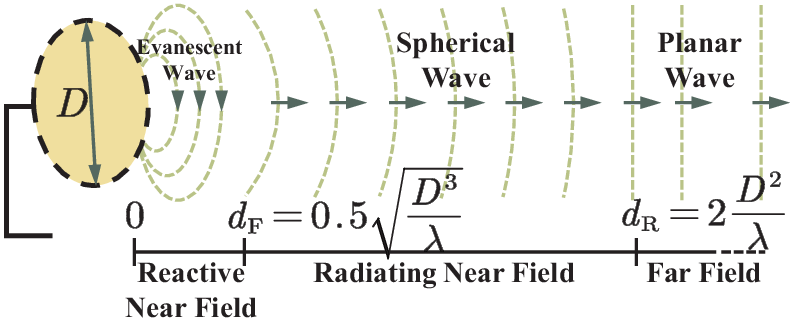}
\caption{Illustration of the EM field.}
\label{Figure: EM_Region}
\vspace{-10pt}
\end{figure}

The electromagnetic (EM) field emitted by an antenna element can be classified into three distinct regions \cite{balanis2016antenna}: the \emph{reactive near-field} region, the \emph{radiating near-field} region, and \emph{the far-field} region, as illustrated in {\figurename} {\ref{Figure: EM_Region}}. The boundary between the reactive and radiating near-field regions is estimated at $d_{\rm{F}}=0.5\sqrt{D^3/\lambda}$, while the boundary between the radiating near-field and far-field regions is estimated at $d_{\rm{R}}=2 D^2/\lambda$, where $D$ represents the antenna aperture and $\lambda$ denotes the wavelength. Future wireless networks are anticipated to utilize extremely large aperture arrays and extremely high frequencies, thereby significantly expanding the previously neglected near-field region of traditional wireless systems \cite{liu2023nearfield}. While far-field EM propagation is commonly approximated using planar waves, near-field propagation necessitates precise modeling using spherical waves. Consequently, the concept of near-field communications (NFC) has gained prominence.

A crucial performance metric for NFC is the channel gain, which quantitatively describes the advantage of large aperture arrays in power transfer. To date, the channel gain has been extensively analyzed in existing literature under both spatially-discrete (SPD) and continuous-aperture (CAP) arrays; see \cite{liu2024near,hu2018beyond,dardari2020communicating,lu2021communicating,liu2023near} and the references therein. However, it is important to note that all existing research has been conducted by disregarding the existence of the reactive near-field region, even when the array aperture size is considered infinite. This oversight may lead to inaccuracies in the results.

As discussed in \cite{liu2023nearfield}, the reactive near-field region of an SPD array is confined to a small vicinity around each antenna element, which typically spans a few wavelengths. However, whether the collective influence of the reactive region of all antenna elements on the total channel gain can be neglected remains uncertain, particularly when the antenna number is infinitely large. Furthermore, in NFC systems employing CAP arrays, the reactive near-field region can significantly expand due to the expansive continuous radiating aperture. In such cases, overlooking the influence of the reactive near field is not straightforward when discussing the near-field channel gain. Within the reactive region, the energy of the EM field oscillates rather than dissipates from the transmitter, which is predominantly characterized by non-propagating evanescent waves (EWs). Predicting the behavior of EWs poses intricate challenges \cite{balanis2016antenna,liu2023nearfield}. To date, the impact of EWs on the channel gain remains unexplored.

To address this research gap, this article analyzes the impact of the reactive region on near-field channel gain. Our main contributions are summarized as follows: {\romannumeral1}) We derive new expressions for the near-field channel gains of both SPD and CAP arrays by considering both radiating and reactive components of the EM field; {\romannumeral2}) For gleaning further insights, we derive closed-form expressions for the asymptotic channel gains in the large limit of the aperture size; {\romannumeral3}) For both SPD and CAP arrays, we prove that the asymptotic near-field channel gain, when considering the reactive region, is smaller compared to its counterpart without such consideration. Besides, we show their gap decays rapidly with the propagation distance; {\romannumeral4}) Through numerical simulations, we illustrate that the impact of the reactive region on near-field channel gain is \emph{negligible}, even as the array aperture size approaches infinity.
\section{System Model}\label{Section: System Model}
\begin{figure}[!t]
    \centering
    \subfigbottomskip=0pt
	\subfigcapskip=-5pt
\setlength{\abovecaptionskip}{0pt}
    \subfigure[SPD array.]
    {
        \includegraphics[height=0.19\textwidth]{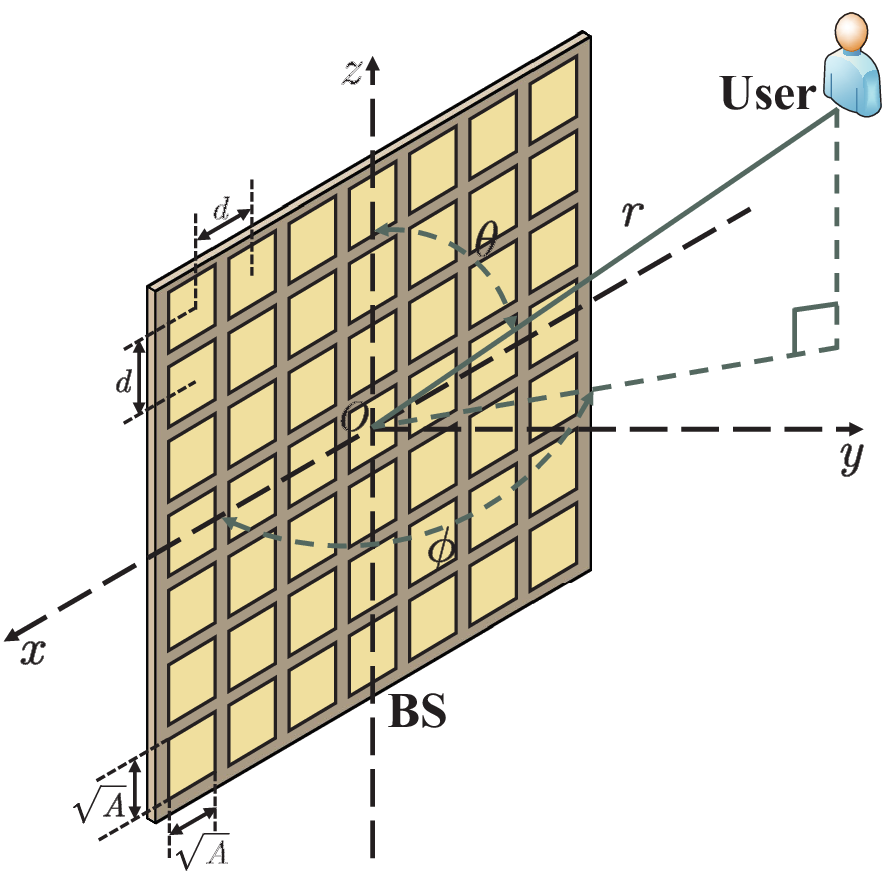}
	   \label{fig1a}	
    }
   \subfigure[CAP array.]
    {
        \includegraphics[height=0.19\textwidth]{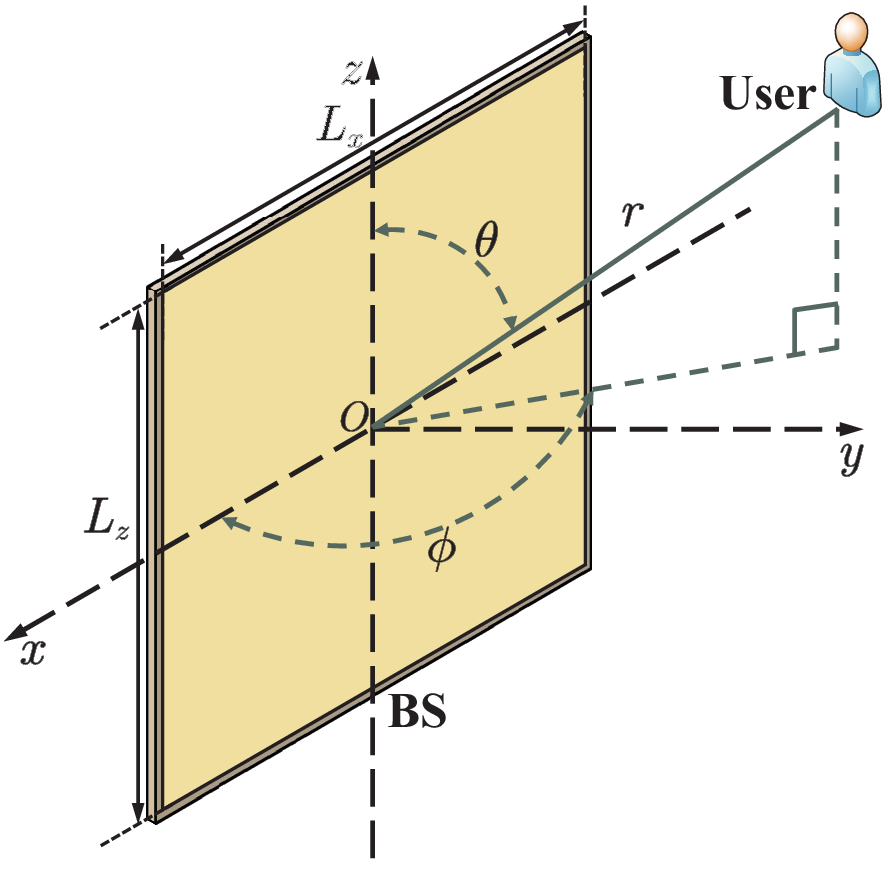}
	   \label{fig1b}	
    }
\caption{Illustration of an SPD array and a CAP array.}
    \label{Figure: System_Model}
    \vspace{-10pt}
\end{figure}

We investigate an uplink channel where a single-antenna user transmits signals to a base station (BS) equipped with an extremely large aperture uniform planar array (UPA), as shown in {\figurename} {\ref{Figure: System_Model}}. The UPA is placed on the $x$-$z$ plane and centered at the origin, with physical dimensions $L_x$ and $L_z$ along the $x$- and $z$-axes, respectively. The user employs a hypothetical isotropic antenna for signal transmission. Let $r$ denote the propagation distance from the center of the antenna array to the user's location, and $\phi\in[0,\pi]$ and $\theta\in[0,\pi]$ denote the associated azimuth and elevation angles, respectively. The user's center location can thus be written as ${\mathbf{s}}_{\mathsf{u}}=[r\Phi,r\Psi,r\Theta]^{\mathsf{T}}$, where $\Phi\triangleq\cos{\phi}\sin{\theta}$, $\Psi\triangleq\sin\phi\sin\theta$, and $\Theta\triangleq\cos{\theta}$.

Let us define ${\mathsf{J}}({\mathbf{s}})\in{\mathbbmss{C}}$ (in amperes per meter, [${\textit{A}}/{\textit{m}}$]) as the continuous distribution of source currents generated by the user, where ${\mathbf{s}}\in{\mathbbmss{R}}^{3\times1}$ represents the source point within the transmit aperture ${\mathcal{A}}_{\mathsf{S}}$. To facilitate analyses, we consider a near-field line-of-sight channel. In this case, the electric radiation field ${\mathsf{E}}(\mathbf{r})\in{\mathbbmss{C}}$ (in volts per meter, [${\textit{V}}/{\textit{m}}$]) generated at point $\mathbf{r}\in{\mathbbmss{R}}^{3\times1}$ within the BS aperture ${\mathcal{A}}_{\mathsf{R}}=\{(x,0,z)|x\in[-\frac{L_x}{2},\frac{L_x}{2}],z\in[-\frac{L_z}{2},\frac{L_z}{2}]\}$ can be mathematically expressed as follows \cite{weinstein1988electromagnetic,balanis2016antenna}:
{\setlength\abovedisplayskip{2pt}
\setlength\belowdisplayskip{2pt}
\begin{equation}\label{General_electric_radiation_field}
{\mathsf{E}}(\mathbf{r})=\int_{{\mathcal{A}}_{\mathsf{S}}}{\mathsf{g}}(\mathbf{r},{\mathbf{s}})
\sqrt{\frac{{\mathbf{e}}_y^{\mathsf{T}}({\mathbf{s}}-{\mathbf{r}})}{\lVert{\mathbf{r}}-{\mathbf{s}}\rVert}}{\mathsf{J}}
({\mathbf{s}}){\rm{d}}{\mathbf{s}}.
\end{equation}
}Here, $\frac{{\mathbf{e}}_y^{\mathsf{T}}({\mathbf{s}}-{\mathbf{r}})}{\lVert{\mathbf{r}}-{\mathbf{s}}\rVert}$ models the impact of the projected aperture of the BS array, which is reflected by the projection of the UPA normal vector ${\mathbf{e}}_y=[0,1,0]^{\mathsf{T}}$ onto the wave propagation direction at each local point $\mathbf{s}$ \cite{liu2023near}. The function ${\mathsf{g}}(\mathbf{r},{\mathbf{s}})$ (in ohms per square meter, [${\it{\Omega}}/{{{\textit{m}}}^2}$]) is given by \cite{dardari2020communicating,bjornson2021primer} 
{\setlength\abovedisplayskip{2pt}
\setlength\belowdisplayskip{2pt}
\begin{equation}\label{Green_Function_Full_Version}
{\mathsf{g}}(\mathbf{r},{\mathbf{s}})=\frac{{\rm{j}}k_0\eta{\rm{e}}^{-{\rm{j}}k_0\lVert{\mathbf{r}}-{\mathbf{s}}\rVert}
}{4\pi \lVert{\mathbf{r}}-{\mathbf{s}}\rVert}
\bigg(1+\frac{{\rm{j}}/k_0}{\lVert{\mathbf{r}}-{\mathbf{s}}\rVert}-\frac{1/k_0^2}{\lVert{\mathbf{r}}-{\mathbf{s}}\rVert^2}\bigg).
\end{equation}
}This function models the influence of free-space EM propagation, where $\eta=120\pi$ (in ohms, [$\it{\Omega}$]) represents the impedance of free space, and $k_0=\frac{2\pi}{\lambda}$ is the wavenumber with $\lambda$ denoting the wavelength. Typically, ${\mathsf{g}}(\mathbf{r},{\mathbf{s}})$ can be treated as the scalar representation of the dyadic Green's function, which comprises three terms: the first term corresponds to the \emph{radiating near-field and far-field} regions, while the remaining two terms correspond to the \emph{reactive near-field} region \cite{poon2005degrees}.

Given the above EM model, the transmitted power (in watts, [${\textit{W}}$]) from the user can be expressed as follows \cite{weinstein1988electromagnetic,balanis2016antenna}:
{\setlength\abovedisplayskip{2pt}
\setlength\belowdisplayskip{2pt}
\begin{equation}\label{SPD_Transmit_Power}
P_{\mathsf{S}}=\int_{{\mathcal{A}}_{\mathsf{S}}}\lvert{\mathsf{J}}
({\mathbf{s}})\rvert^2 {\mathsf{R}}_{\mathsf{rad}}{\rm{d}}{\mathbf{s}},
\end{equation}
}where ${\mathsf{R}}_{\mathsf{rad}}$ (in ohms, [$\it{\Omega}$]) denotes the radiation resistance. The total power received by the BS is given by \cite{weinstein1988electromagnetic,balanis2016antenna}
{\setlength\abovedisplayskip{2pt}
\setlength\belowdisplayskip{2pt}
\begin{equation}\label{SPD_Receive_Power}
P_{\mathsf{R}}=\int_{{\mathcal{S}}_{\mathsf{R}}}\frac{1}{2\eta}\lvert{\mathsf{E}}
({\mathbf{r}})\rvert^2 {\rm{d}}{\mathbf{r}},
\end{equation}
}where ${\mathcal{S}}_{\mathsf{R}}\subseteq{\mathcal{A}}_{\mathsf{R}}$ represents the effective surface region within the BS aperture that can capture the radiated power. The \emph{channel gain} is thus calculated as ${P_{\mathsf{R}}}/{P_{\mathsf{S}}}$ \cite{liu2023near}. 

In conventional research, the last two reactive terms in ${\mathsf{g}}(\mathbf{r},{\mathbf{s}})$ are often neglected \cite{liu2023near}. By contrast, we retain these terms to examine the impact of the reactive near-field region, viz. EWs, on the channel gain when the array aperture size tends to infinity. For a comprehensive analysis, we consider both SPD and CAP arrays, as depicted in {\figurename} {\ref{Figure: System_Model}}. Throughout this paper, we assume that the aperture size of ${\mathcal{A}}_{\mathsf{R}}$ significantly exceeds that of ${\mathcal{A}}_{\mathsf{S}}$, i.e., $\lvert{\mathcal{A}}_{\mathsf{R}}\rvert\gg \lvert{\mathcal{A}}_{\mathsf{S}}\rvert$. 
\subsection{SPD Arrays}
We begin by considering the scenario in which the BS is equipped with an SPD array comprising $M$ antennas, as shown in {\figurename} {\ref{fig1a}}. Here, $M=M_{x}M_{z}$, where $M_{x}$ and $M_{z}$ denote the number of antenna elements along the $x$- and $z$-axes, respectively. Without loss of generality, we assume that $M_{x}$ and $M_{z}$ are odd numbers, which satisfy $M_x=2\tilde{M}_x+1$ and $M_z=2\tilde{M}_z+1$. 

The physical dimensions of each antenna element along the $x$- and $z$-axes are indicated by $\sqrt{A}$, and the inter-element distance is denoted as $d$, where $d\geq\sqrt{A}$. The central location of the $(m_x,m_z)$th element is given by ${\mathbf{r}}_{m_x,m_z}=[m_xd,0,m_zd]^{\mathsf{T}}$, where $m_x\in{\mathcal{M}}_x\triangleq\{0,\pm1,\ldots,\pm\tilde{M}_x\}$ and $m_z\in{\mathcal{M}}_z\triangleq\{0,\pm1,\ldots,\pm\tilde{M}_z\}$. Thus, the physical dimensions of the UPA satisfy $L_x\approx M_xd$ and $L_z\approx M_zd$. The distance between the center of the user and the center of the $(m_x,m_z)$th antenna element is given by
{\setlength\abovedisplayskip{2pt}
\setlength\belowdisplayskip{2pt}
\begin{equation}\label{General_Distance_Expression}
r_{m_x,m_z}=r((m_x\varepsilon-\Phi)^2+{\Psi}^2+(m_z\varepsilon-\Theta)^2)^{\frac{1}{2}},
\end{equation}
}where $\varepsilon=\frac{d}{r}$. Note that $r=r_{0,0}$ and, since the antenna element separation $d$ is typically on the order of a wavelength, in practice, we have $\varepsilon\ll1$. Under the SPD configuration, the effective surface area ${\mathcal{S}}_{\mathsf{R}}$ is defined as the sum over all $M=M_{x}M_{z}$ antenna elements:
{\setlength\abovedisplayskip{2pt}
\setlength\belowdisplayskip{2pt}
\begin{equation}
{\mathcal{S}}_{\mathsf{R}}=\bigcup\nolimits_{m_x\in{\mathcal{M}}_x,m_z\in{\mathcal{M}}_z}{\mathcal{S}}_{m_x,m_z},
\end{equation}
}where ${\mathcal{S}}_{m_x,m_z}=\{(m_xd+\ell,0,m_zd+\ell)|\ell\in[-\frac{\sqrt{A}}{2},\frac{\sqrt{A}}{2}]\}$ represents the surface region of the $(m_x,m_z)$th antenna element. Thus, the total power received by the BS is given by
{\setlength\abovedisplayskip{2pt}
\setlength\belowdisplayskip{2pt}
\begin{equation}\label{SPD_Receive_Power}
P_{\mathsf{R}}=\sum\nolimits_{m_x=-\tilde{M}_x}^{\tilde{M}_x}\sum\nolimits_{m_z=-\tilde{M}_z}^{\tilde{M}_z}
\int_{{\mathcal{S}}_{m_x,m_z}}\frac{1}{2\eta}\lvert{\mathsf{E}}
({\mathbf{r}})\rvert^2 {\rm{d}}{\mathbf{r}}.
\end{equation}
}
\subsection{CAP Arrays} 
We are also interested in the channel gain when the BS is equipped with a CAP array, as depicted in {\figurename} {\ref{fig1b}}. A CAP array can be regarded as an SPD array constructed using an infinite number of antennas with infinitesimal spacing. In this context, the effective surface region is the entire aperture of the CAP array, which yields ${\mathcal{S}}_{\mathsf{R}}={\mathcal{A}}_{\mathsf{R}}$. Consequently, the total power received by the BS can be expressed as follows:
{\setlength\abovedisplayskip{2pt}
\setlength\belowdisplayskip{2pt}
\begin{equation}\label{CAP_Receive_Power}
P_{\mathsf{R}}=
\int_{{\mathcal{A}}_{\mathsf{R}}}\frac{1}{2\eta}\lvert{\mathsf{E}}
({\mathbf{r}})\rvert^2 {\rm{d}}{\mathbf{r}},
\end{equation}
}
\section{SPD Arrays}
\subsection{Analysis of the Channel Gain}
Given that the size of the transmit aperture $\lvert{\mathcal{A}}_{\mathsf{S}}\rvert$ is considerably smaller than the receive aperture $\lvert{\mathcal{A}}_{\mathsf{R}}\rvert$, the transmit power in \eqref{SPD_Transmit_Power} can be simplified as follows \cite{liu2023near}:
{\setlength\abovedisplayskip{2pt}
\setlength\belowdisplayskip{2pt}
\begin{equation}\label{Appr_Transmit_Power}
P_{\mathsf{S}}\approx\lvert{\mathsf{J}}
({\mathbf{s}}_{\mathsf{u}})\rvert^2 \lvert{\mathcal{A}}_{\mathsf{S}}\rvert {\mathsf{R}}_{\mathsf{rad}},
\end{equation}
}and the electric radiation field in \eqref{General_electric_radiation_field} satisfies
{\setlength\abovedisplayskip{2pt}
\setlength\belowdisplayskip{2pt}
\begin{equation}\label{General_electric_radiation_field_appr}
{\mathsf{E}}(\mathbf{r})\approx {\mathsf{g}}(\mathbf{r},{\mathbf{s}}_{\mathsf{u}})
\sqrt{\frac{{\mathbf{e}}_y^{\mathsf{T}}({\mathbf{s}}_{\mathsf{u}}-{\mathbf{r}})}{\lVert{\mathbf{r}}-{\mathbf{s}}_{\mathsf{u}}
\rVert}}{\mathsf{J}}({\mathbf{s}}_{\mathsf{u}})\lvert{\mathcal{A}}_{\mathsf{S}}\rvert.
\end{equation}
}Besides, due to the small antenna size, $\sqrt{A}$, compared to the distance between the user and antenna elements $\lVert{\mathbf{r}}_{m_x,m_z}-{\mathbf{s}}_{\mathsf{u}}\rVert$, the variation of the electric radiation field within an antenna element is negligible \cite{liu2023near}. This leads to
{\setlength\abovedisplayskip{2pt}
\setlength\belowdisplayskip{2pt}
\begin{align}\label{SPD_Electric_Field_Antenna_Element}
\int_{{\mathcal{S}}_{m_x,m_z}}\lvert{\mathsf{E}}
({\mathbf{r}})\rvert^2 {\rm{d}}{\mathbf{r}}\approx \lvert{\mathsf{E}}({\mathbf{r}}_{m_x,m_z})\rvert^2 A.
\end{align}
}Inserting \eqref{Green_Function_Full_Version}, \eqref{General_electric_radiation_field_appr}, and \eqref{SPD_Electric_Field_Antenna_Element} into \eqref{SPD_Receive_Power} gives
{\setlength\abovedisplayskip{2pt}
\setlength\belowdisplayskip{2pt}
\begin{equation}
\begin{split}
P_{\mathsf{R}}&\approx\frac{\lvert{\mathsf{J}}({\mathbf{s}}_{\mathsf{u}})\rvert^2\lvert{\mathcal{A}}_{\mathsf{S}}\rvert^2 k_0^2\eta^2}{(2\eta)(4\pi)^2}
\sum\nolimits_{m_x=-\tilde{M}_x}^{\tilde{M}_x}\sum\nolimits_{m_z=-\tilde{M}_z}^{\tilde{M}_z}\\
&\times {Ar\Psi}
\frac{\Big\lvert1+\frac{{\rm{j}}/k_0}{\lVert{\mathbf{r}}_{m_x,m_z}-{\mathbf{s}}_{\mathsf{u}}\rVert}-
\frac{1/k_0^2}{\lVert{\mathbf{r}}_{m_x,m_z}-{\mathbf{s}}_{\mathsf{u}}\rVert^2}\Big\rvert^2}
{\lVert{\mathbf{r}}_{m_x,m_z}-{\mathbf{s}}_{\mathsf{u}}\rVert^3},
\end{split}
\end{equation}
}where $\lVert{\mathbf{r}}_{m_x,m_z}-{\mathbf{s}}_{\mathsf{u}}\rVert=r_{m_x,m_z}$ is calculated using \eqref{General_Distance_Expression}.

For clarity, we define the following function:
{\setlength\abovedisplayskip{2pt}
\setlength\belowdisplayskip{2pt}
\begin{align}
f_n(x,z)\triangleq((x-\Phi)^2+{\Psi}^2+(z-\Theta)^2)^{-\frac{n}{2}}.
\end{align}
}Accordingly, the receive power can be rewritten as follows:
{\setlength\abovedisplayskip{2pt}
\setlength\belowdisplayskip{2pt}
\begin{equation}\label{SPD_Power_Gain_Summation}
\begin{split}
P_{\mathsf{R}}&\approx\frac{\lvert{\mathsf{J}}({\mathbf{s}}_{\mathsf{u}})\rvert^2\lvert{\mathcal{A}}_{\mathsf{S}}\rvert^2 k_0^2\eta^2}{(2\eta)(4\pi)^2}
\sum_{m_x,m_z}\frac{A\Psi \varepsilon^2}{d^2}\Big(f_3(m_x\varepsilon,m_z\varepsilon)\\
&-\frac{1}{k_0^2 r^2}f_5(m_x\varepsilon,m_z\varepsilon)
+\frac{1}{k_0^4 r^4}f_7(m_x\varepsilon,m_z\varepsilon)\Big).
\end{split}
\end{equation}
}Since the user is equipped with a hypothetical isotropic antenna, we have $\lvert{\mathcal{A}}_{\mathsf{S}}\rvert=\frac{\lambda^2}{4\pi}$ \cite{balanis2016antenna}. By combining this with \eqref{SPD_Power_Gain_Summation}, we derive the channel gain as follows.
\vspace{-5pt}
\begin{theorem}\label{SPD_Power_Gain_Explicit_Theorem}
When the BS is equipped with an SPD array, the channel gain is given by
{\setlength\abovedisplayskip{2pt}
\setlength\belowdisplayskip{2pt}
\begin{equation}\label{SPD_Power_Gain_Integral}
\begin{split}
\frac{P_{\mathsf{R}}}{P_{\mathsf{S}}}&\approx\frac{\eta}{8
{\mathsf{R}}_{\mathsf{rad}}}\frac{A\Psi}{4\pi d^2}\iint_{\mathcal{H}}\Big(f_3(x,z)\\
&-\frac{1}{k_0^2 r^2}f_5(x,z)+\frac{1}{k_0^4 r^4}f_7(x,z)\Big){\rm{d}}x{\rm{d}}z\triangleq {\mathsf{G}}_{\mathsf{p}},
\end{split}
\end{equation}
}where ${\mathcal{H}}=\{\left.(x,z)\right|-\frac{M_x\varepsilon}{2}\leq x\leq\frac{M_x\varepsilon}{2},-\frac{M_z\varepsilon}{2}\leq z\leq\frac{M_z\varepsilon}{2}\}$.
\end{theorem}
\vspace{-5pt}
\begin{IEEEproof}
The rectangular area ${\mathcal{H}}$ can be partitioned into $M_xM_z$ sub-rectangles, each with equal area $\varepsilon^2$. Since $\varepsilon\ll 1$, we have $f_n(x,z)\approx f_n(m_x\varepsilon,m_z\varepsilon)$ for $\forall (x,z)\in\{\left.(x,z)\right|(m_x-\frac{1}{2})\varepsilon\leq x\leq(m_x+\frac{1}{2})\varepsilon,(m_z-\frac{1}{2})\varepsilon\leq z\leq(m_z+\frac{1}{2})\varepsilon\}$. By the concept of double integrals, we have
{\setlength\abovedisplayskip{2pt}
\setlength\belowdisplayskip{2pt}
\begin{align}\label{SPD_Power_Gain_Integral_Fundamental}
\sum_{m_x\in{\mathcal{M}}_x}\sum_{m_z\in{\mathcal{M}}_z}f_n(m_x\varepsilon,m_z\varepsilon)\varepsilon^2\approx
\iint_{\mathcal{H}}f_n(x,z){\rm{d}}x{\rm{d}}z.
\end{align}
}By inserting \eqref{SPD_Power_Gain_Integral_Fundamental} and $\lvert{\mathcal{A}}_{\mathsf{S}}\rvert=\frac{\lambda^2}{4\pi}$ into \eqref{SPD_Power_Gain_Summation} as well as calculating ${P_{\mathsf{R}}}/{P_{\mathsf{S}}}$, we obtain \eqref{SPD_Power_Gain_Integral}. 
\end{IEEEproof}
\subsection{Impact of the Reactive Region}
Equation \eqref{SPD_Power_Gain_Integral} can be solved in closed-form but the final expression is quite articulated and it does not provide important insights. Therefore, for the sake of space, we report here the result valid for $M_x,M_z\rightarrow\infty$ from which the impact of the reactive region on the channel gain can be unveiled. In this case, the following theorem can be derived.
\vspace{-5pt}
\begin{theorem}\label{SPD_Power_Gain_Asym_Theorem}
When the BS is equipped with an SPD UPA, the channel gain satisfies
{\setlength\abovedisplayskip{2pt}
\setlength\belowdisplayskip{2pt}
\begin{equation}\label{SPD_Power_Gain_Asym_UPA_Final}
\begin{split}
\lim_{M_x,M_z\rightarrow\infty}{\mathsf{G}}_{\mathsf{p}}=
\frac{\eta \mu_{\mathsf{oc}}}{8
{\mathsf{R}}_{\mathsf{rad}}}\Big(\frac{1}{2}-\frac{\frac{\lambda^2}{r^2\Psi^2}}{24\pi^2}+\frac{\frac{\lambda^4}{r^4\Psi^4}}
{160\pi^4}\Big)\triangleq {\mathsf{G}}_{\mathsf{eva}}^{\mathsf{upa}},
\end{split}
\end{equation}
}where $\mu_{\mathsf{oc}}\triangleq\frac{A}{d^2}\leq 1$ represents the array occupation ratio.
\end{theorem}
\vspace{-5pt}
\begin{IEEEproof}
Please refer to Appendix for more details.
\end{IEEEproof}
By following similar steps to derive \eqref{SPD_Power_Gain_Asym_UPA_Final}, we obtain the channel gain in the absence of EWs, i.e., by omitting the last two terms in \eqref{Green_Function_Full_Version}, as follows:
{\setlength\abovedisplayskip{2pt}
\setlength\belowdisplayskip{2pt}
\begin{equation}\label{SPD_Power_Gain_Asym_UPA_Normal}
\begin{split}
\lim_{M_x,M_z\rightarrow\infty}{\mathsf{G}}_{\mathsf{p}}=
\frac{\eta \mu_{\mathsf{oc}}}{8  {\mathsf{R}}_{\mathsf{rad}}}\frac{1}{2}\triangleq {\mathsf{G}}_{\mathsf{rad}}^{\mathsf{upa}}.
\end{split}
\end{equation}
}This expression was derived under the assumption that $\frac{\eta}{8 {\mathsf{R}}_{\mathsf{rad}}}=1$, which is termed as the normalized channel gain \cite{liu2024near,hu2018beyond,dardari2020communicating,lu2021communicating,liu2023near}. 
\vspace{-5pt}
\begin{remark}
Upon comparing ${\mathsf{G}}_{\mathsf{eva}}^{\mathsf{upa}}$ with ${\mathsf{G}}_{\mathsf{rad}}^{\mathsf{upa}}$, we observe that considering the reactive region adds two reactive terms to ${\mathsf{G}}_{\mathsf{rad}}^{\mathsf{upa}}$, both related to the perpendicular distance from the user to the BS array, i.e., $r\Psi$. In other words, the influence of EWs on the asymptotic channel gain for SPD UPAs introduces additional range dimensions to ${\mathsf{G}}_{\mathsf{rad}}^{\mathsf{upa}}$.
\end{remark}
\vspace{-5pt}
For comparison, we also present the channel gain under the far-field model, where variations in channel amplitude $\lVert{\mathbf{r}}-{\mathbf{s}}_{\mathsf{u}}\rVert$ along the BS array are considered negligible, which yields 
{\setlength\abovedisplayskip{2pt}
\setlength\belowdisplayskip{2pt}
\begin{equation}
\lvert{\mathsf{E}}
({\mathbf{r}})\rvert^2\approx \frac{k_0^2\eta^2\Psi}{(4\pi)^2r^2}
\lvert{\mathsf{J}}({\mathbf{s}}_{\mathsf{u}})\rvert^2\lvert{\mathcal{A}}_{\mathsf{S}}\rvert^2;
\end{equation}
}therefore, the channel gain can be expressed as follows:
{\setlength\abovedisplayskip{2pt}
\setlength\belowdisplayskip{2pt}
\begin{equation}\label{SPD_Power_Gain_Asym_FFC_Final}
\begin{split}
\frac{P_{\mathsf{R}}}{P_{\mathsf{S}}}\approx \frac{\eta}{8 {\mathsf{R}}_{\mathsf{rad}}}\frac{A\Psi}{4\pi r^2}M\triangleq {\mathsf{G}}_{\mathsf{far}}^{\mathsf{upa}},
\end{split}
\end{equation}
}which scales linearly with the antenna number $M$.
\vspace{-5pt}
\begin{remark}\label{Far-Field_Discussion}
The linear growth trend in \eqref{SPD_Power_Gain_Asym_FFC_Final} has the potential to elevate the receive power to arbitrary levels, even exceeding the transmit power, thereby violating the law of conservation of energy. This phenomenon arises because, as $M$ increases, the planar-wave propagation-based far-field model cannot capture the exact physical properties of near-field propagation.
\end{remark}
\vspace{-5pt}
Combining \eqref{SPD_Power_Gain_Asym_UPA_Final} with \eqref{SPD_Power_Gain_Asym_UPA_Normal} gives
{\setlength\abovedisplayskip{2pt}
\setlength\belowdisplayskip{2pt}
\begin{equation}\label{SPD_Array_Power_Gain}
\frac{{\mathsf{G}}_{\mathsf{eva}}^{\mathsf{upa}}}{{\mathsf{G}}_{\mathsf{rad}}^{\mathsf{upa}}}=
1-\frac{1}{12\pi^2}\frac{\lambda^2}{r^2\Psi^2}+\frac{1}{80\pi^4}\frac{\lambda^4}{r^4\Psi^4}\triangleq {\mathsf{r}}_{\mathsf{spd}},
\end{equation}
}the value of which is less than $1$ when $r\Psi > ({\frac{3}{20}})^{\frac{1}{2}}\frac{\lambda}{\pi}\approx 0.12\lambda$. This implies that EWs can degrade the channel gain, viz. ${{\mathsf{G}}_{\mathsf{eva}}^{\mathsf{upa}}}<{{\mathsf{G}}_{\mathsf{rad}}^{\mathsf{upa}}}$, in practical scenarios where $r\gg \lambda$.
\vspace{-5pt}
\begin{remark}\label{remark_EWs_Nagative}
The above result aligns with intuition, as EWs lack the ability to contribute effectively to propagating waves or radiating energy. Within the reactive near-field region, the energy of the EM field oscillates rather than dissipates from the transmitter. Thus, considering the reactive region tends to yield a smaller channel gain compared to neglecting it.
\end{remark}
\vspace{-5pt}
To delve deeper, let us study the degeneration level of the reactive region on the channel gain. Assuming $r\Psi=\lambda$ gives
{\setlength\abovedisplayskip{2pt}
\setlength\belowdisplayskip{2pt}
\begin{equation}\label{Degenaration_Power_Gain_SPD}
\begin{split}
{\mathsf{r}}_{\mathsf{spd}}=
1-\frac{1}{12\pi^2}+\frac{1}{80\pi^4}\approx 99.17\%.
\end{split}
\end{equation}
}This value is very close to $1$. Additionally, we note that both $\frac{1}{12\pi^2}\frac{\lambda^2}{r^2\Psi^2}$ and $\frac{1}{80\pi^4}\frac{\lambda^4}{r^4\Psi^4}$ decay rapidly with the propagation distance $r$. Thus, in practical scenarios where $r\gg \lambda$, the disparity between ${{\mathsf{G}}_{\mathsf{eva}}^{\mathsf{upa}}}$ and ${{\mathsf{G}}_{\mathsf{rad}}^{\mathsf{upa}}}$ is negligible. Consequently, considering the reactive region has minimal impact on the overall channel gain in general system configurations. 
\vspace{-5pt}
\begin{remark}\label{EW_Influence_SPD}
These observations suggest that the impact of the reactive region on the channel gain can be safely disregarded, even in the case of an infinitely large SPD planar array.
\end{remark}
\vspace{-5pt} 
\subsection{Uniform Linear Array}\label{Subsection: Uniform Linear Array}
We then explore the scenario where the BS is equipped with a uniform linear array (ULA) to gain further insights into the impact of the reactive region on the channel gain. Specifically, we assume that $M_z=1$ and $M_x=M>1$. Under this setup, the double integral in \eqref{SPD_Power_Gain_Integral} simplifies into a single integral, which can be solved in closed-form.
\vspace{-5pt}
\begin{corollary}
When the BS is equipped with an SPD ULA, the channel gain is given by
{\setlength\abovedisplayskip{2pt}
\setlength\belowdisplayskip{2pt}
\begin{equation}\label{SPD_Power_Gain_ULA_Calculation}
\begin{split}
&\frac{P_{\mathsf{R}}}{P_{\mathsf{S}}}\approx\frac{\eta}{32\pi{\mathsf{R}}_{\mathsf{rad}}}
\frac{A\Psi}{dr}\sum\nolimits_{x\in\{\frac{M\epsilon}{2}\pm\Phi\}}\bigg(\frac{x}{a_\Phi^2\sqrt{x^2+a_\Phi^2}}\\
&-\frac{\frac{1}{k_0^2 r^2}x(3a_\Phi^2+2x^2)}{3 a_\Phi^4(a_\Phi^2+x^2)^{\frac{3}{2}}}+
\frac{x(15a_\Phi^4+20a_\Phi^2x^2+8x^4)}{15 k_0^4 r^4 a_\Phi^6(a_\Phi^2+x^2)^{\frac{5}{2}}}\bigg),
\end{split}
\end{equation}
}where $a_\Phi^2=1-\Phi^2$. When $M\rightarrow\infty$, ${P_{\mathsf{R}}}/{P_{\mathsf{S}}}$ satisfies
{\setlength\abovedisplayskip{2pt}
\setlength\belowdisplayskip{2pt}
\begin{equation}\label{SPD_Power_Gain_ULA_Asym_Calculation}
\begin{split}
\lim_{M\rightarrow\infty}\frac{P_{\mathsf{R}}}{P_{\mathsf{S}}}\approx
\frac{\eta A\Psi\left(1-\frac{{2}/{3}}{k_0^2 r^2 a_\Phi^2}+
\frac{{8}/{15}}{k_0^4 r^4 a_\Phi^4}\right)}{16\pi{\mathsf{R}}_{\mathsf{rad}}d r a_\Phi^2}\triangleq
{\mathsf{G}}_{\mathsf{eva}}^{\mathsf{ula}}.
\end{split}
\end{equation}
}\end{corollary}
\vspace{-5pt}
\begin{IEEEproof}
This corollary is derived by setting $m_z$ in \eqref{SPD_Power_Gain_Summation} as $m_z=0$ and using the concept of single integrals. Then \eqref{SPD_Power_Gain_ULA_Asym_Calculation} is obtained by calculating the resultant integral.
\end{IEEEproof}
By omitting the last two terms within the parenthesis in \eqref{SPD_Power_Gain_ULA_Asym_Calculation}, we obtain the asymptotic channel gain that only considers the radiating region, which is given by
{\setlength\abovedisplayskip{2pt}
\setlength\belowdisplayskip{2pt}
\begin{equation}\label{SPD_Radiating_Power_Gain_ULA_Asym_Calculation}
\begin{split}
\lim_{M\rightarrow\infty}\frac{P_{\mathsf{R}}}{P_{\mathsf{S}}}\approx
\frac{\eta A\Psi}{16\pi{\mathsf{R}}_{\mathsf{rad}}d r a_\Phi^2}\triangleq
{\mathsf{G}}_{\mathsf{rad}}^{\mathsf{ula}},
\end{split}
\end{equation}
}which is related to the user's distance to the ULA.
\vspace{-5pt}
\begin{remark}
In contrast to the UPA case, the reactive region under the ULA scenario does not introduce additional range dimensions to the asymptotic channel gain. 
\end{remark}
\vspace{-5pt}
It is evident that for general system settings with $r\gg \lambda$, we have ${\mathsf{G}}_{\mathsf{rad}}^{\mathsf{ula}}>{\mathsf{G}}_{\mathsf{eva}}^{\mathsf{ula}}$. Moreover, we observe that ${{\mathsf{G}}_{\mathsf{eva}}^{\mathsf{ula}}}/{{\mathsf{G}}_{\mathsf{rad}}^{\mathsf{ula}}}$ equals $98.35\%$ at distance $ra_{\Phi}=\lambda$.
\vspace{-5pt}
\begin{remark}
These observations suggest that for SPD ULAs, the impact of the reactive region on the asymptotic channel gain lies in introducing additional degenerated terms. However, similar to the UPA case, the corresponding degeneration level can generally be neglected.
\end{remark}
\vspace{-5pt}
\section{CAP Arrays}\label{Section: CAP Arrays}
In this section, we explore the scenario where the BS is equipped with a CAP array, i.e., ${\mathcal{S}}_{\mathsf{R}}={\mathcal{A}}_{\mathsf{R}}$. Following the derivation steps outlined in obtaining Theorem \ref{SPD_Power_Gain_Explicit_Theorem}, we calculate the channel gain for the CAP array as follows.
\vspace{-5pt}
\begin{theorem}\label{CAP_Power_Gain_Asym_Theorem}
When the BS is equipped with a CAP array, the channel gain is given by
{\setlength\abovedisplayskip{2pt}
\setlength\belowdisplayskip{2pt}
\begin{equation}\label{SPD_Power_Gain_Integral1}
\begin{split}
&\frac{P_{\mathsf{R}}}{P_{\mathsf{S}}}\approx\frac{\eta}{8
{\mathsf{R}}_{\mathsf{rad}}}\frac{\Psi}{4\pi}\iint_{\tilde{\mathcal{H}}}\Big(f_3(x,z)\\
&-\frac{1}{k_0^2 r^2}f_5(x,z)+\frac{1}{k_0^4 r^4}f_7(x,z)\Big){\rm{d}}x{\rm{d}}z\triangleq \tilde{\mathsf{G}}_{\mathsf{p}}.
\end{split}
\end{equation}
}where $\tilde{\mathcal{H}}=\{\left.(x,z)\right|-\frac{L_x}{2r}\leq x\leq\frac{L_x}{2r},-\frac{L_z}{2r}\leq z\leq\frac{L_z}{2r}\}$.
\end{theorem}
\vspace{-5pt}
\begin{IEEEproof}
Similar to the proof of Theorem \ref{SPD_Power_Gain_Explicit_Theorem}.
\end{IEEEproof}
Using the approximation $L_x\approx M_xd$ and $L_z\approx M_z d$, we can establish the following corollary.
\vspace{-5pt}
\begin{corollary}\label{corollary_power_gain_ratio}
The channel gain ratio satisfies ${{\mathsf{G}}_{\mathsf{p}}}/{\tilde{\mathsf{G}}_{\mathsf{p}}}\approx \mu_{\mathsf{oc}}$.
\end{corollary}
\vspace{-5pt}
\begin{IEEEproof}
Since $L_x\approx M_xd$ and $L_z\approx M_z d$, we have ${\mathcal{H}}\approx \tilde{\mathcal{H}}$, and the final result follows immediately.
\end{IEEEproof}
\vspace{-5pt}
\begin{remark}\label{SPD_CAP_Relation}
The result in Corollary \ref{corollary_power_gain_ratio} suggests that \eqref{SPD_Power_Gain_Integral1} can be treated as a special case of \eqref{SPD_Power_Gain_Integral} by setting the array occupation ratio as $\mu_{\mathsf{oc}}=1$. This observation makes intuitive sense as a CAP array is a special case of an SPD array when the array occupation ratio equals one.
\end{remark}
\vspace{-5pt}
In the sequel, we investigate the asymptotic channel gain for CAP arrays by letting $L_x,L_z\rightarrow\infty$.
\vspace{-5pt}
\begin{corollary}\label{CAP_Power_Gain_Asym_Corollary}
When the BS is equipped with a CAP array, the channel gain satisfies
{\setlength\abovedisplayskip{2pt}
\setlength\belowdisplayskip{2pt}
\begin{equation}\label{CAP_Power_Gain_Asym_UPA_Final}
\begin{split}
\lim_{L_x,L_z\rightarrow\infty}\tilde{\mathsf{G}}_{\mathsf{p}}=
\frac{\eta }{8
{\mathsf{R}}_{\mathsf{rad}}}\left(\frac{1}{2}-\frac{\frac{\lambda^2}{r^2\Psi^2}}{24\pi^2}+\frac{\frac{\lambda^4}{r^4\Psi^4}}{160\pi^4}
\right)\triangleq \tilde{\mathsf{G}}_{\mathsf{eva}}^{\mathsf{upa}}.
\end{split}
\end{equation}
}When omitting the reactive terms, we have
{\setlength\abovedisplayskip{2pt}
\setlength\belowdisplayskip{2pt}
\begin{equation}\label{CAP_Power_Gain_Asym_UPA_Normal}
\begin{split}
\lim_{L_x,L_z\rightarrow\infty}\tilde{\mathsf{G}}_{\mathsf{p}}=
\frac{\eta}{8  {\mathsf{R}}_{\mathsf{rad}}}\frac{1}{2}\triangleq \tilde{\mathsf{G}}_{\mathsf{rad}}^{\mathsf{upa}}.
\end{split}
\end{equation}
}\end{corollary}
\vspace{-5pt}
\begin{IEEEproof}
Similar to the derivation steps of \eqref{SPD_Power_Gain_Asym_UPA_Final} and \eqref{SPD_Power_Gain_Asym_UPA_Normal}.
\end{IEEEproof}
It follows that ${\tilde{\mathsf{G}}_{\mathsf{eva}}^{\mathsf{upa}}}/{\tilde{\mathsf{G}}_{\mathsf{rad}}^{\mathsf{upa}}}=
{\mathsf{r}}_{\mathsf{spd}}$, which is the same as its SPD counterpart. Thus, the insights derived from channel gain limits in SPD arrays readily extend to CAP arrays. Moreover, by following the derivation steps outlined in Section \ref{Subsection: Uniform Linear Array}, we can also discuss the channel gain when $L_z$ is fixed and $L_x$ approaches infinity. Further elaboration on this aspect is omitted for brevity.
\section{Numerical Results}
In this section, the impact of the reactive region on the near-field channel gain is studied through computer simulation results. Unless otherwise specified, we fix the parameters as follows: $\frac{\eta}{8 {\mathsf{R}}_{\mathsf{rad}}}=1$, $\theta=\frac{\pi}{6}$, $\phi=\frac{\pi}{3}$, $r=5$ m, $d = 0.0628$ m, $\lambda=2d$, $A=\frac{\lambda^2}{4\pi}$, $M_x=M_z$, and $L_x=L_z$. For the sake of brevity, all simulations use planar arrays.
\begin{figure}[!t]
    \centering
    \subfigbottomskip=0pt
	\subfigcapskip=-5pt
\setlength{\abovecaptionskip}{0pt}
    \subfigure[SPD.]
    {
        \includegraphics[height=0.175\textwidth]{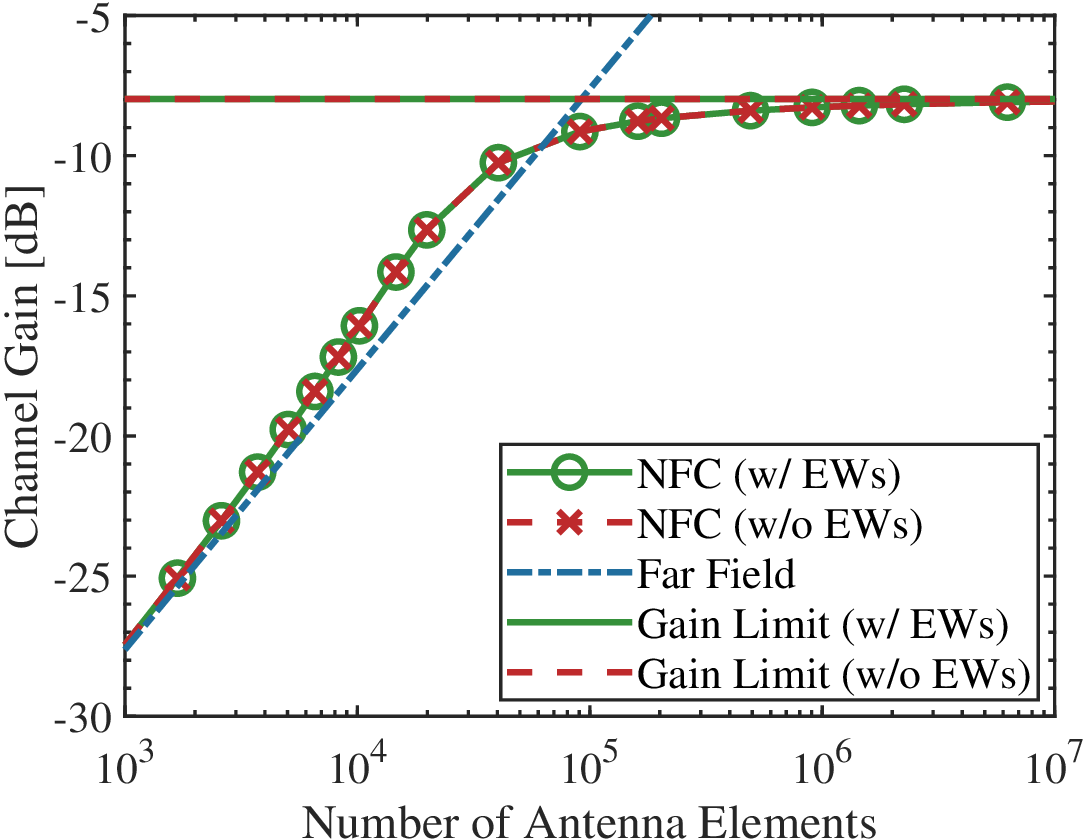}
	   \label{fig2a}	
    }
   \subfigure[CAP.]
    {
        \includegraphics[height=0.175\textwidth]{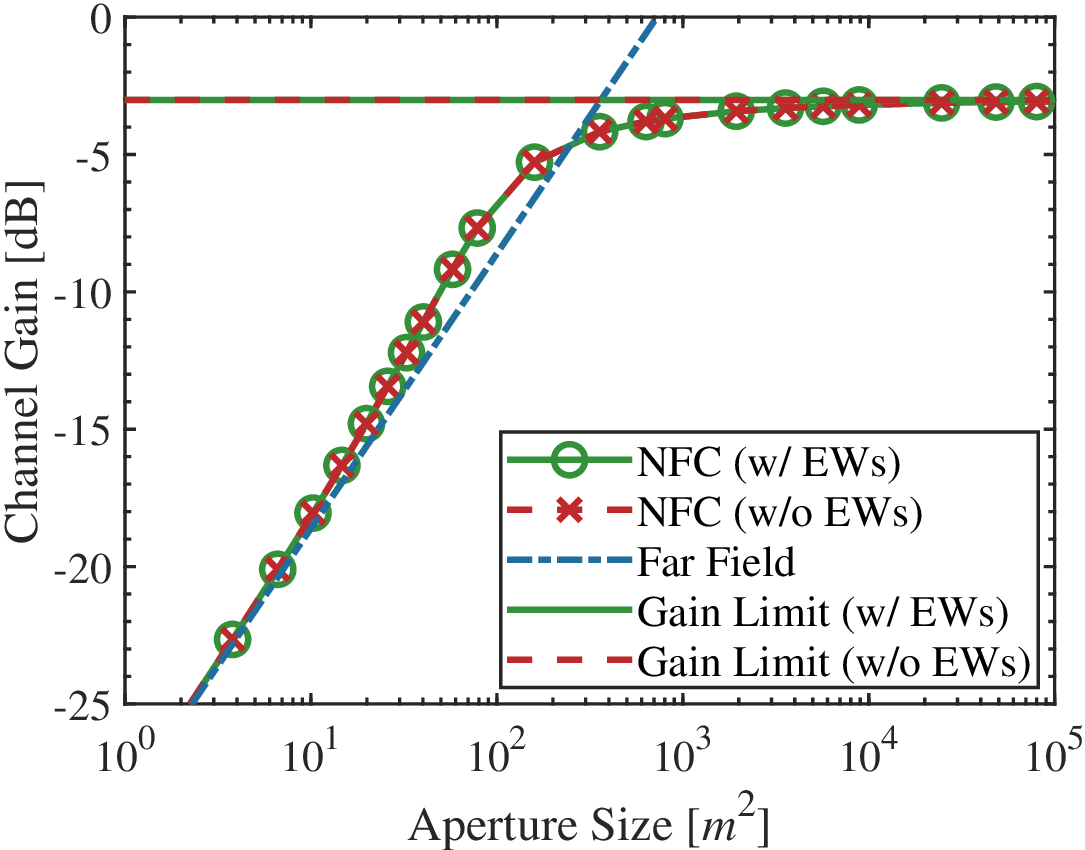}
	   \label{fig2b}	
    }
\caption{Channel gains of SPD and CAP arrays.}
    \label{Figure2}
    \vspace{-15pt}
\end{figure}

{\figurename} {\ref{fig2a}} and {\figurename} {\ref{fig2b}} illustrate the channel gain of the SPD array and the CAP array, respectively. As observed, as the number of antennas ($M$) or the aperture size ($\lvert{\mathcal{A}}_{\mathsf{R}}\rvert$) increases, the simulated channel gain converges to its limit, which thus validates the analytical results presented in Theorem \ref{SPD_Power_Gain_Asym_Theorem} and Corollary \ref{CAP_Power_Gain_Asym_Corollary}. Upon comparison, the CAP array exhibits a higher channel gain compared to the SPD array, which is in line with the conclusion in Remark \ref{SPD_CAP_Relation}. Furthermore, for sufficiently large values of $M$ or $\lvert{\mathcal{A}}_{\mathsf{R}}\rvert$, the far-field channel gain exhibits a linear increase, which violates the principle of energy conservation. This occurs because the user resides in the near field as $M\rightarrow\infty$ or $\lvert{\mathcal{A}}_{\mathsf{R}}\rvert\rightarrow\infty$, where the far-field model inadequately characterizes EM propagations, as discussed in Remark \ref{Far-Field_Discussion}. In contrast, the near-field channel gain converges to a finite constant limit.

Our focus then shifts to evaluating the impact of the reactive region or EWs on the channel gain. As depicted in {\figurename} {\ref{Figure2}}, for both SPD and CAP arrays, the near-field channel gain considering EWs closely aligns with that without considering EWs, even as $M$ or $\lvert{\mathcal{A}}_{\mathsf{R}}\rvert$ tends to infinity. This validation reinforces the points outlined in Remark \ref{EW_Influence_SPD} and highlights the negligible impact of the reactive region on the channel gain, even with an infinitely large array aperture. 
\begin{figure}[!t]
 \centering
\setlength{\abovecaptionskip}{0pt}
\includegraphics[height=0.25\textwidth]{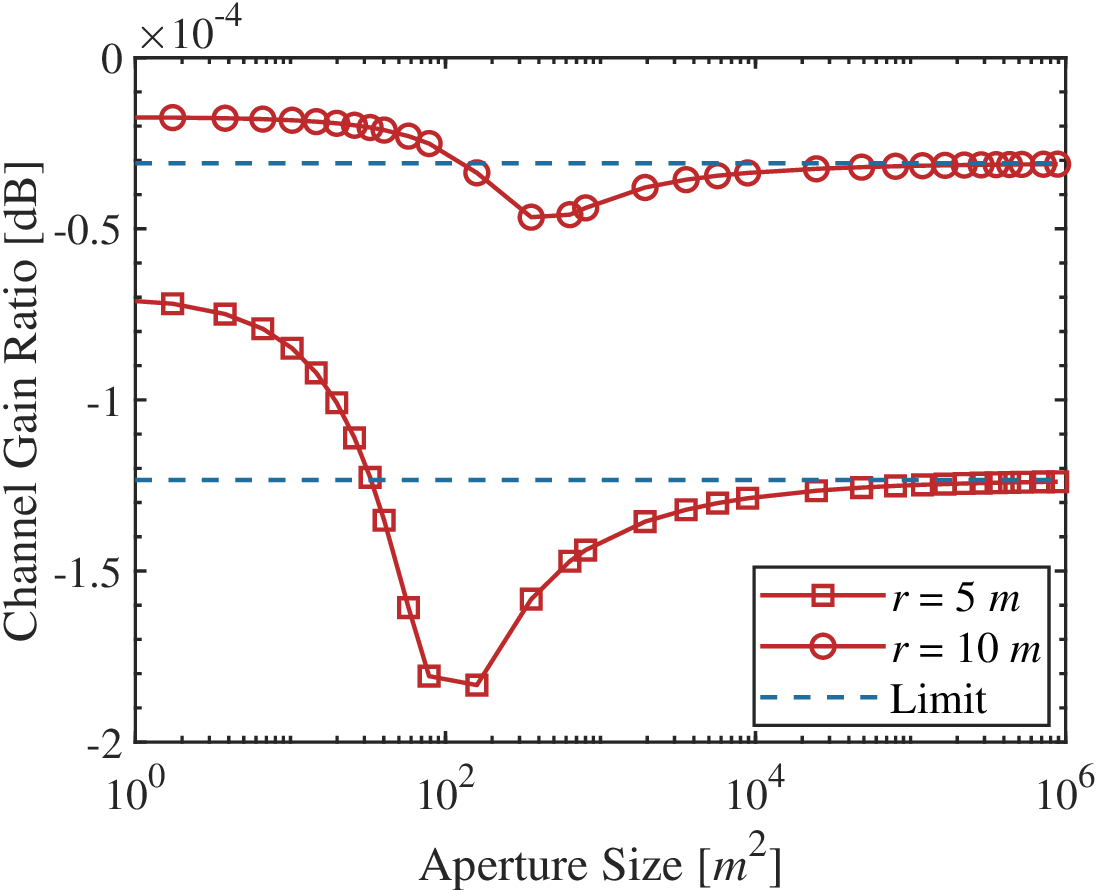}
\caption{Channel gain ratio.}
\label{Figure3}
\vspace{-15pt}
\end{figure}

After discussing the channel gain, we proceed to illustrate in {\figurename} {\ref{Figure3}} the ratio between the channel gain with and without considering the reactive region. Since ${{\mathsf{G}}_{\mathsf{p}}}/{\tilde{\mathsf{G}}_{\mathsf{p}}}\approx \mu_{\mathsf{oc}}$, the channel gain ratio remains the same for both SPD and CAP arrays. Therefore, we only depict the channel gain ratio for CAP arrays in {\figurename} {\ref{Figure3}}, which is calculated as $\frac{\iint_{\tilde{\mathcal{H}}}(f_3(x,z)-\frac{1}{k_0^2 r^2}f_5(x,z)+\frac{1}{k_0^4 r^4}f_7(x,z)){\rm{d}}x{\rm{d}}z}{\iint_{\tilde{\mathcal{H}}}f_3(x,z){\rm{d}}x{\rm{d}}z}$. As depicted, the gain ratio gradually converges to its limit ${\tilde{\mathsf{G}}_{\mathsf{eva}}^{\mathsf{upa}}}/{\tilde{\mathsf{G}}_{\mathsf{rad}}^{\mathsf{upa}}}=
{\mathsf{r}}_{\mathsf{spd}}$ as $\lvert{\mathcal{A}}_{\mathsf{R}}\rvert$ increases. Besides, the gain ratio decreases with the propagation distance $r$. For all aperture sizes, the gain ratio remains below $0$ dB, which indicates that the channel gain with the reactive region considered is lower than that without considering it, thereby confirming the validity of Remark \ref{remark_EWs_Nagative}. However, it is noteworthy that for all considered aperture size ranges, the channel gain ratio nearly approaches $0$ dB, which suggests that the degradation level of the reactive region on the channel gain can be neglected. 
\section{Conclusion} 
\begin{figure}[!t]
 \centering
\setlength{\abovecaptionskip}{0pt}
\includegraphics[height=0.25\textwidth]{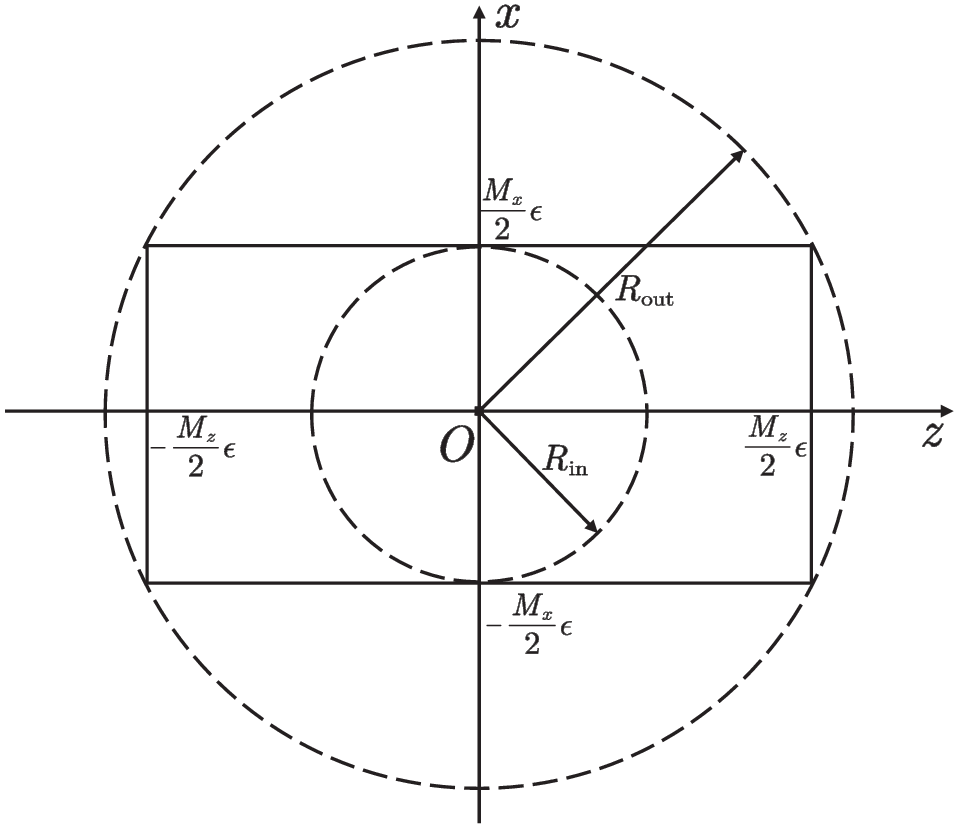}
\caption{The inscribed and circumscribed disks of the rectangular region ${\mathcal{H}}=\{\left.(x,z)\right|\frac{M_x\varepsilon}{-2}\leq x\leq\frac{M_x\varepsilon}{2},\frac{M_z\varepsilon}{-2}\leq z\leq\frac{M_z\varepsilon}{2}\}$.}
\label{Figure: Geometry}
\vspace{-15pt}
\end{figure}

We have derived novel expressions for the near-field channel gains of SPD and CAP arrays by taking into account both radiating and reactive components of the EM field. By further assuming an infinitely large aperture size, we have demonstrated that the difference between channel gains with and without consideration of the reactive region diminishes rapidly with increasing propagation distance. This diminishment makes the impact of the reactive region negligible, even as the array aperture size tends towards infinity. As a result, we conclude that in practical near-field analyses, the presence of the reactive region can be disregarded for both SPD and CAP arrays.
\begin{appendix}
The integral $\iint_{\mathcal{H}}f_n(x,z){\rm{d}}x{\rm{d}}z$ can be rewritten as follows:
{\setlength\abovedisplayskip{2pt}
\setlength\belowdisplayskip{2pt}
\begin{equation}
\begin{split}
\int_{-\frac{M_z\varepsilon}{2}-\Theta}^{\frac{M_z\varepsilon}{2}-\Theta}
\int_{-\frac{M_x\varepsilon}{2}-\Phi}^{\frac{M_x\varepsilon}{2}-\Phi}
\frac{1}{(x^2+z^2+\Psi^2)^{\frac{n}{2}}}{\rm{d}}x{\rm{d}}z.
\end{split}
\end{equation}
}Since $\Phi\in[0,1]$ and $\Theta\in[0,1]$, we have
{\setlength\abovedisplayskip{2pt}
\setlength\belowdisplayskip{2pt}
\begin{equation}\label{Received_SNR_Analog_SU_Near_Field_Trans3}
\begin{split}
\lim_{M_x,M_z\rightarrow\infty}\frac{\iint_{\mathcal{H}}f_n(x,z){\rm{d}}x{\rm{d}}z}
{\int_{-\frac{M_z\varepsilon}{2}}^{\frac{M_z\varepsilon}{2}}\int_{-\frac{M_x\varepsilon}{2}}^{\frac{M_x\varepsilon}{2}}
\frac{1}{(x^2+z^2+\Psi^2)^{\frac{n}{2}}}{\rm{d}}x{\rm{d}}z}=1.
\end{split}
\end{equation}
}As shown in {\figurename} {\ref{Figure: Geometry}}, the integration region in the denominator of \eqref{Received_SNR_Analog_SU_Near_Field_Trans3} is bounded by two disks with radius $R_{\mathsf{in}}\triangleq\frac{\varepsilon}{2}\min\{M_x,M_z\}$ and $R_{\mathsf{out}}\triangleq\frac{\varepsilon}{2}\sqrt{M_x^2+M_z^2}$, respectively. 
By defining function
{\setlength\abovedisplayskip{2pt}
\setlength\belowdisplayskip{2pt}
\begin{align}
\hat{f}_n(R)&\triangleq\int_{0}^{2\pi}\int_{0}^{R}\frac{\rho }{(\rho^2+{\Psi}^2)^{\frac{n}{2}}}{\rm{d}}\rho {\rm{d}}\vartheta\\
&={\pi}({{n}/{2}-1})^{-1}({\Psi}^{2-n}-(\Psi^2+R^2)^{1-\frac{n}{2}}),
\end{align}
}we have
{\setlength\abovedisplayskip{2pt}
\setlength\belowdisplayskip{2pt}
\begin{equation}\label{Double_Integral_Bound}
\hat{f}_n(R_{\mathsf{in}}) < \iint_{\mathcal{H}}\frac{{\rm{d}}x{\rm{d}}z}{(x^2+z^2+\Psi^2)^{\frac{n}{2}}} < \hat{f}_n(R_{\mathsf{out}}).
\end{equation}
}As $M_x,M_z\rightarrow\infty$, we have $R_{\mathsf{in}},R_{\mathsf{out}}\rightarrow\infty$, which, together with the definitions of $R_{\mathsf{in}}$ and $R_{\mathsf{out}}$, yields
{\setlength\abovedisplayskip{2pt}
\setlength\belowdisplayskip{2pt}
\begin{equation}\label{Double_Integral_Bound_Limitataion}
\lim_{M_x,M_z\rightarrow\infty}{\hat{f}_n(R_{\mathsf{in}})}=\lim_{M_x,M_z\rightarrow\infty}{\hat{f}_n(R_{\mathsf{out}})}
=\frac{\pi{\Psi}^{2-n}}{{n}/{2}-1}
\end{equation}
}for $n=3,5,7$. Using the results in \eqref{Received_SNR_Analog_SU_Near_Field_Trans3}, \eqref{Double_Integral_Bound}, and \eqref{Double_Integral_Bound_Limitataion} as well as the sandwich theorem, we obtain
{\setlength\abovedisplayskip{2pt}
\setlength\belowdisplayskip{2pt}
\begin{equation}
\lim_{M_x,M_z\rightarrow\infty}\iint_{\mathcal{H}}\frac{{\rm{d}}x{\rm{d}}z}{(x^2+z^2+\Psi^2)^{\frac{n}{2}}}
=\frac{\pi{\Psi}^{2-n}}{{n}/{2}-1},
\end{equation}
}which, together with \eqref{Received_SNR_Analog_SU_Near_Field_Trans3}, yields
{\setlength\abovedisplayskip{2pt}
\setlength\belowdisplayskip{2pt}
\begin{equation}\label{Double_Integral_Limitataion}
\lim_{M_x,M_z\rightarrow\infty}\iint_{\mathcal{H}}f_n(x,z){\rm{d}}x{\rm{d}}z=\frac{\pi{\Psi}^{2-n}}{{n}/{2}-1}.
\end{equation}
}Inserting \eqref{Double_Integral_Limitataion} into \eqref{SPD_Power_Gain_Integral} gives \eqref{SPD_Power_Gain_Asym_UPA_Final}. Theorem \ref{SPD_Power_Gain_Asym_Theorem} is thus proved.
\end{appendix}
\bibliographystyle{IEEEtran}
\bibliography{mybib}
\end{document}